\documentclass[11pt,twoside]{article}

\usepackage{asp2021}

\aspSuppressVolSlug
\resetcounters

\begin{document}

\title{The Vera C. Rubin Observatory Prompt Processing System}

\author{Krzysztof~Findeisen,$^1$ Kian\nobreakdashes-Tat~Lim,$^2$ Dan~Speck,$^3$ Hsin\nobreakdashes-Fang~Chiang,$^2$ Erin~Leigh~Howard,$^1$ Ian~S.~Sullivan,$^1$ and Eric~C.~Bellm$^1$
\affil{$^1$University of Washington, Seattle, WA, USA; \email{kfindeis@uw.edu}}
\affil{$^2$SLAC National Accelerator Laboratory, Menlo~Park, CA, USA}
\affil{$^3$Burwood Group, Oak~Brook, IL, USA}}

\paperauthor{Krzysztof~Findeisen}{kfindeis@uw.edu}{0000-0003-1898-5760}{University of Washington}{Dept.\ of Astronomy}{Seattle}{WA}{98195}{USA}
\paperauthor{Kian\nobreakdashes-Tat~Lim}{ktl@slac.stanford.edu}{0000-0002-6338-6516}{SLAC National Accelerator Laboratory}{}{Menlo~Park}{CA}{94025}{USA}
\paperauthor{Dan~Speck}{dspeck@burwood.com}{}{Burwood Group}{}{Oak~Brook}{IL}{60523}{USA}
\paperauthor{Hsin\nobreakdashes-Fang~Chiang}{hfc@stanford.edu}{0000-0002-1181-1621}{SLAC National Accelerator Laboratory}{}{Menlo~Park}{CA}{94025}{USA}
\paperauthor{Erin~Leigh~Howard}{elhoward@uw.edu}{0000-0002-0716-947X}{University of Washington}{Dept.\ of Astronomy}{Seattle}{WA}{98195}{USA}
\paperauthor{Ian~S.~Sullivan}{sullii@uw.edu}{0000-0001-8708-251X}{University of Washington}{Dept.\ of Astronomy}{Seattle}{WA}{98195}{USA}
\paperauthor{Eric~C.~Bellm}{ecbellm@uw.edu}{0000-0001-8018-5348}{University of Washington}{Dept.\ of Astronomy}{Seattle}{WA}{98195}{USA}



\begin{abstract}
Vera C. Rubin Observatory's Prompt Processing system will automatically process 10~TB of raw images to produce up to 10~million transient alerts per night.
We summarize how Prompt Processing meets its throughput, latency, and reliability requirements
and present results from Rubin Observatory Commissioning.
\end{abstract}



\section{Introduction}
\label{130-intro}

The NSF-DOE Vera C. Rubin Observatory's Legacy Survey of Space and Time \citep[LSST,][]{2019ApJ...873..111I} will generate unprecedented data volumes.
The LSST Camera \citep{2024SPIE13096E..1SR}, with 189 4K$\times$4K detectors, produces 8~GB of uncompressed data every 39~s and 10~TB per night.
LSST data will be processed in two modes: annual Data Releases and real-time Prompt Processing.
Prompt Processing handles images immediately, identifying new, variable, or moving sources and generating up to 10,000 alerts per visit and 10~million per night, all within 60-120~s of shutter close.

This paper describes how the Prompt Processing system (at \url{https://github.com/lsst-dm/prompt_processing/}) handles a large data volume to produce alerts with low latency, with little downtime or human intervention, while executing the Alert Production pipeline \citep{PSTN-019}.

\section{Prompt Processing Design}
\label{130-design}

\subsection{Key Components}
\label{130-design-arch}

To flexibly scale with changes to observing cadence, processing time, and other variables, \citet{DMTN-219} proposed a task-based architecture with one task for each detector of each visit (one ``visit-detector'').
The tasks are submitted to a worker pool running on a Kubernetes cluster, each worker processing its task independently in an individual workspace.
Independent execution maximizes performance, per Amdahl's law, while a cloud-native worker pool benefits from existing solutions for autoscaling, error recovery, metrics, etc.

While \citet{DMTN-219} originally implemented Prompt Processing with the Knative Serverless Platform, we migrated to Kubernetes Event-Driven Autoscaling (KEDA) in early 2025 after finding Knative is ill-suited for long-running tasks.
The current implementation can scale up to 1700 Kubernetes pods, limited by cluster memory.
Each pod handles multiple tasks over its lifetime, amortizing startup costs and allowing caching between runs.
A shared Redis Streams queue tracks tasks and holds any overflow.

\articlefigure[width=0.93\linewidth]{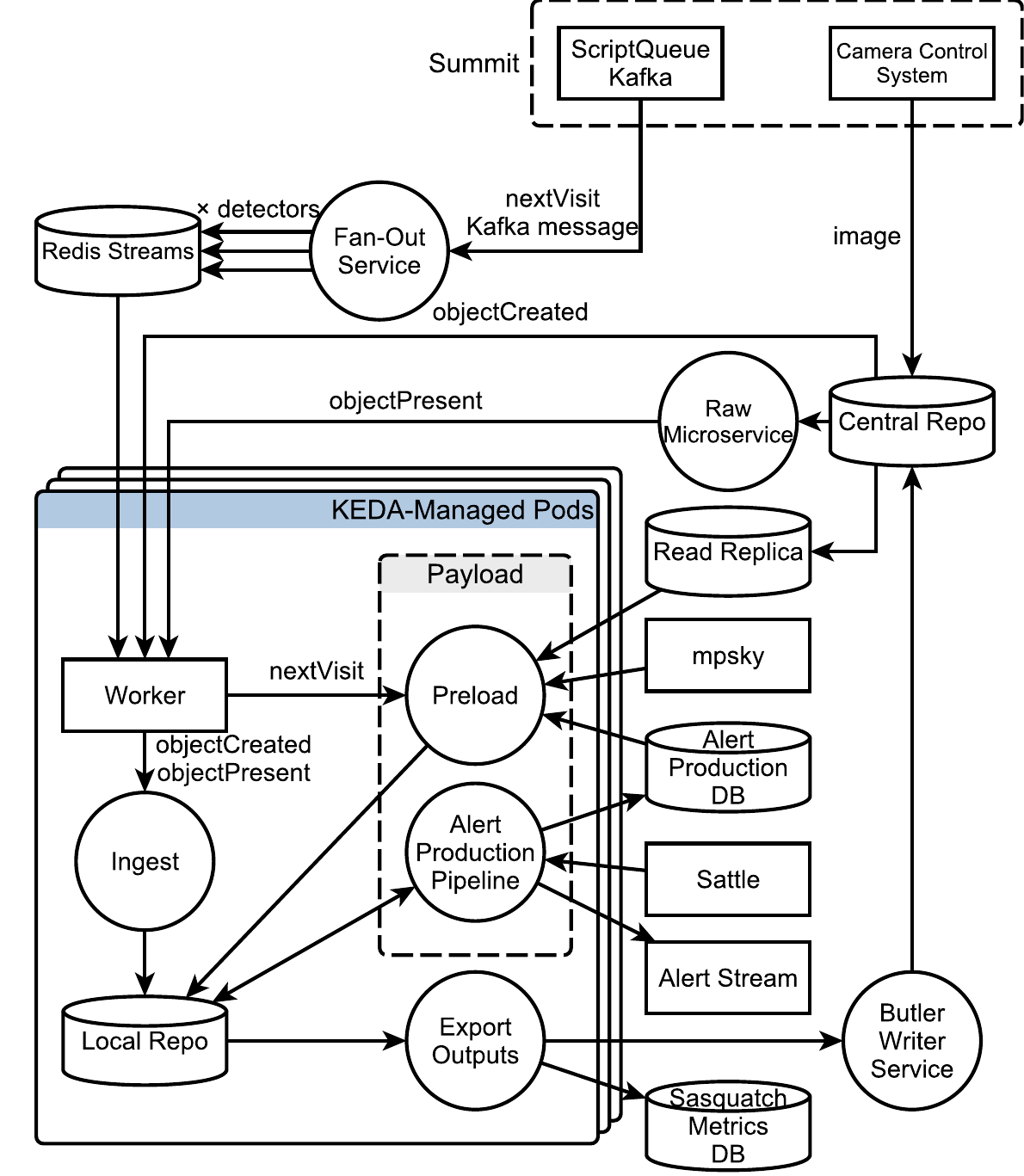}{130-fig-dataflow}{Data flow diagram for the Prompt Processing system, showing interaction with nextVisit messages and other Rubin Observatory systems.}

While the system confines processing to independent workers as much as possible, it must interact with several shared resources, shown in Fig.~\ref{130-fig-dataflow}: a central data repository \citep{2022SPIE12189E..11J} to load calibrations, difference imaging templates, and other inputs, and to store pipeline products; the Cassandra Alert Production database \citep[APDB,][]{DMTN-293} to load and update a record of the variable sky; the \texttt{mpsky} ephemeris service to identify solar system objects; and the Sattle service to vet candidate sources against known satellites.

\subsection{Workflow}
\label{130-design-workflow}

To minimize alert latency, visit processing begins before the image is taken.
As shown in Fig.~\ref{130-fig-dataflow}, the Summit facility sends a ``nextVisit'' message one visit or 20~s (whichever is longer) before exposure start, which a microservice converts into visit-detector tasks.
On picking up a new task, each worker preloads the needed inputs from the central repository, the APDB, and \texttt{mpsky}.
Preload normally finishes before the image arrives, letting pipeline execution begin immediately.
Workers cache calibration data, but this is not practical for template images and other field-specific inputs.

Normally, image arrival triggers a Kafka notification that the worker picks up with minimal delay.
As a backup, a microservice indexes previously received images.
When either mechanism detects the awaited image, the worker ingests it into its own workspace and runs the pipeline locally.
Whether the pipeline succeeds or fails, the worker exports pipeline products and metrics as a separate step.

Prompt Processing does not have a hard-coded pipeline, instead abstracting pipe\-lines through the Rubin Observatory Middleware framework \citep{2022SPIE12189E..11J}.
This lets the system configure different pipelines based on nextVisit metadata, as well as prioritized lists of pipelines to be attempted.

As noted in Section~\ref{130-design-arch}, avoiding shared resources or operations is essential to good parallel performance.
External I/O is minimized, and restricted as much as possible to preload and export, where any delays contribute to worker load but not alert latency.
There are three exceptions: the Alert Production pipeline writes to the APDB and the alert stream, which are both optimized for bulk writes, and queries the Sattle service, which is low latency.
Sending alerts before cleaning up pipeline execution not only gets them to alert brokers faster, but also minimizes the risk that the alerts will be lost if another operation fails \citep{DMTN-260}.

Pipeline failures are handled through the Middleware framework, and by default assumed to be repeatable.
If a worker fails for non-algorithmic reasons, such as I/O errors or pod termination, we consider a visit-detector retriable if it has not updated the APDB \citep{DMTN-260}.
Such retries are not yet implemented with KEDA.

\section{Commissioning Results}
\label{130-commissioning}

Prompt Processing was first deployed in May~2023, handling live imaging data from the LATISS instrument on the Rubin Auxiliary Telescope \citep{10.71929/rubin/2571930}.
It ran at increasing scale when the LSST Commissioning Camera \citep{2022SPIE12184E..0JS} and the LSST Camera \citep{2024SPIE13096E..1SR} went on sky in November~2024 and April~2025, respectively.
This two and a half year commissioning process gave us experience running Prompt Processing in a variety of failure scenarios.

Prompt Processing sent 4~million alerts in 1\nobreakdashes-2~hours each night on July~1 and July~2 to Rubin's alert brokers, demonstrating the required throughput.
However, the central repository and the APDB struggled to handle over a thousand simultaneous clients.
Slow or timed out queries degraded latency and reliability.
The Middleware team added a read-only database replica and a serialized writer service, while the Alert Production and Data Access teams streamlined APDB read and write volume.
These will be tested at scale as survey-like observations resume.

From July~12 to the end of LSSTCam Commissioning, Prompt Processing handled 73\% of all science images, with 22\% lost when the read replica fell out of date (blocking data access until manually reset), 3\% when unexpected camera rotations invalidated preload, and 2\% to other causes.
These issues have since been mitigated: the read replica is more reliable, and has redundant checks and alerts to detect desynchronization earlier, while scheduler and Summit workflow improvements will ensure that images are almost always taken as originally planned.

\section{Ongoing Work}
\label{130-concl}

We continue to identify and mitigate failure modes during the Rubin Early Optimization Phase \citep{RTN-011}, and plan to restore support for retrying transient errors in KEDA.
We also need to test existing solutions to the shared-resource bottlenecks at full survey cadence, though smaller tests are encouraging.
Reducing latency will benefit more from optimizing pipeline payloads than Prompt Processing itself.

There will inevitably be failures requiring human intervention, particularly if recovery involves reconciling APDB history.
We plan to create a daytime processing service to handle this small number of cases.

\acknowledgements This material is based upon work supported in part by the National Science Foundation through Cooperative Agreements AST-1258333 and AST-2241526 and Cooperative Support Agreements AST-1202910 and 2211468 managed by the Association of Universities for Research in Astronomy (AURA), and the Department of Energy under Contract No. DE-AC02-76SF00515 with the SLAC National Accelerator Laboratory managed by Stanford University. Additional Rubin Observatory funding comes from private donations, grants to universities, and in-kind support from LSST-DA Institutional Members.

\bibliography{130}

\end{document}